\documentclass[amsmath,amssymb]{revtex4}

\usepackage{graphicx}
\usepackage{dcolumn}
\usepackage{bm}
\usepackage{epsfig}
\usepackage{slashed}
\usepackage{subfigure}

\begin{document}

\def\0#1#2{\frac{#1}{#2}}
\def\bct{\begin{center}} \def\ect{\end{center}}
\def\beq{\begin{equation}} \def\eeq{\end{equation}}
\def\bea{\begin{eqnarray}} \def\eea{\end{eqnarray}}
\def\nnu{\nonumber}
\def\n{\noindent} \def\pl{\partial}
\def\g{\gamma}  \def\O{\Omega} \def\e{\varepsilon} \def\o{\omega}
\def\s{\sigma}  \def\b{\beta} \def\p{\psi} \def\r{\rho}
\def\G{\Gamma} \def\k{\kappa} \def\l{\lambda} \def\d{\delta}

\title{Vacuum and thermal fluctuation energies of a soliton at finite temperatures}
\author{Song Shu$^1$}
\email {shus@hubu.edu.cn}
\author{Xiaogang Li$^1$}
\author{Jia-Rong Li$^{2}$}

 \affiliation{1. Department of Physics and Electronic Science, Hubei
University, Wuhan, Hubei 430062, China}
 \affiliation{2. Key Laboratory of Quark and Lepton Physics (MOE) and Institute of Particle Physics, Central China Normal University, Wuhan, Hubei 430079, China}

\begin{abstract}
We have studied the vacuum and thermal fluctuation energies of a soliton at finite temperatures.
First the vacuum energy coming from the Dirac sea is calculated by
a summation of the discrete and continuum energy
spectrum of the Dirac equation in the background field of a soliton. And all the divergences are removed
by the same renormalization scheme at zero temperature. Then the vacuum energy and thermal fluctuation energy at finite temperatures in a temperature dependent soliton background are calculated. And the numerical results are analyzed and discussed.
\end{abstract}
\maketitle

\section{Introduction}
In recent years the nontrivial properties of the vacuum, like the Casimir effect in the vacuum of the QED system\cite{ref1,ref2,ref3}, sphaleron and baryogenesis in the vacuum of the electroweak or cosmological system\cite{ref4,ref5}, and especially instanton, chiral symmetry and topology in the QCD vacuum\cite{ref6,ref7,ref8,ref9}, have been studied more seriously by many researchers. As the symmetry breaking generates the condensation from the vacuum which induces the observable energy and mass, in electroweak sector of the Standard Model it triggers the masses of the elementary particles and in strong sector, namely QCD, it generates the confinement and chiral symmetry breaking at low energy scale. Nucleons are the basic tangible blocks of our real world. How does the nucleon mass emerge from the vacuum? As nucleons are built on quarks, the quarks are more fundamental ingredients. In electroweak theory the quark acquires a small initial mass due to the Higgs mechanism. In QCD theory the chiral symmetry is spontaneously broken by nonzero chiral condensate through the nontrivial QCD vacuum. The nontrivial QCD vacuum plays the crucial roles in generating the nucleon mass.

In the early years before QCD was fully established, different kinds of effective quark models had been used to describe nucleons. On one hand there are bag models, such as MIT bag model, SLAC bag model and Friedberg-Lee (FL) model, which are built based on confinement mechanism\cite{ref10,ref11,ref12}. In these bag models the valence quarks are confined in a small spatial area (less than $1 fm$) by some confining potentials or by the surface tension generated by some nontrivial scalar fields. On the other hand there are chiral models like Skyrme model which is linked to QCD at large $N_c$ limit\cite{ref13,ref14}. The nucleon is described by the topological solitons called Skyrmions. It should be noted that there are no valence quarks in this model. Chiral symmetry breaking and topology play crucial roles in this scenario. Based on the chiral symmetry breaking there are also other chiral models to describe the nucleon, such as chiral bag model\cite{ref15,ref16}, Nambu-Jona-Lasinio(NJL) soliton model\cite{ref17}, chiral quark soliton model and etc.\cite{ref18,ref19}. In these models the valence quarks are coupled to the chiral meson fields. The nucleons are depicted as the valence quarks surrounded by the chiral pion cloud. It is indicated that these chiral soliton models could interpolate the valence quark model and the Skyrme model\cite{ref19a}. The ``magic" here is the vacuum polarization. In the valence quark model the vacuum is polarized very slightly that the valence quarks have very large masses in vacuum and are bounded by introducing some confinement mechanism, while in the Skyrme model the vacuum is polarized so much that the lowest positive energy orbit of the Dirac energy spectrum has crossed the surface of the vacuum and got close to the negative continuum. The whole positive energy levels are vacant. The negative energy spectrum has been distorted so remarkably that the baryon number is totally generated by the vacuum polarization. Thus the nucleon is emerged as the result of the vacuum polarization. However the reality, which is indicated by Diakonov\cite{ref20}, maybe stays in the between, that means the soliton meson fields create some discrete bound states above the vacuum in which the lowest positive energy orbital is occupied by the three valence quarks, meanwhile the vacuum is also polarized in a certain degree. The baryon charge of the nucleon is carried by the three valence quarks, while the total energy or mass of the nucleon comes from both the valence quarks and the polarization energy of the vacuum. In this case besides the valence quarks it is also important to study the vacuum energy of the nucleon.

The vacuum energy in chiral quark soliton model can be viewed as the energy of the quantum fluctuation of quarks in a static background of a soliton. However generally speaking the calculation of the vacuum energy in the nonhomogeneous or spatially non-trivial background is nonlocal and quite difficult. In early years the quantum fluctuations namely the Casmimir energy of the electron in the background of a static configuration of electromagnetic fields had been studied by Schwinger\cite{ref21}. The one loop quantum fluctuations were calculated by a scheme of the energy level summation over both discrete and continuous energy spectrum. In particular the summation over the continuum was a nontrivial integration over energy or momentum in which the density of states for the continuum modes was determined by the scattering phase shift. The divergences had been regulated in this scheme. In later years the calculation scheme were extended by Jaffe and his collaborators\cite{ref2,ref22,ref22a}. They further completed the renormalizaton program and developed practical computational methods for calculating the quantum fluctuations of the soliton background in electroweak systems\cite{ref22,ref22a}. From a more general basis of quantum mechanics and field theories a new version of semiclassical theory has been developed recently for calculating the quantum and thermal fluctuations of a particular soliton background called \textit{flucton} which configuration in the Euclidean time has fixed coinciding endpoints\cite{ref22b}.

The practical level summation calculation scheme based on the scattering phase shift is suitable to calculate the vacuum energy of the nucleon through soliton models. In our previous study we have applied this scheme to the FL model in which we have performed the renormalization and numerically evaluated the scattering phase shift\cite{ref23}. The quantum corrections of the soliton in FL model is obtained in zero temperature case. In this paper we want to extend the work to the finite temperature case which means the soliton embedded in a thermal quark medium at finite temperatures. As we know the FL model is a very simple effective model which lacks the chiral symmetry, it is not fully justified to study the nucleon despite its successful phenomenological description of the nucleon. However it is a suitable model for methodology of developing the calculation scheme.

In 1990's it was an important problem to study the hadron properties in medium. In theory the prevailing view was the Brown-Rho scaling which indicated that the hadrons will get lighter when chiral symmetry gets restored or partially restored in hot or dense medium\cite{ref23a}. However things get more complicated when strongly coupled quark-gluon plasma(sQGP) has been found and more significances of nonperturbative and nontrivial sides of QCD vacuum have been revealed\cite{ref23b,ref23c,ref23d,ref23e}. In the future experiments of electron-ion collider(EIC) the composition and structure of the nucleon will be focused on more precisely\cite{ref23f}. It provides a good opportunity to study the quarks and especially the gluons in a nonperturbative regime of the nucleon embedded in the nuclei. In theoretic side it is more important to consider the nontrivial and nonperturbative quantum effect of the gluon fields and the QCD vacuum and develop corresponding calculation scheme in quantum field theory. In present work we mainly focused on making the calculation scheme based on the phase shift method more precise and efficient in studying quantum fluctuations including vacuum and thermal fluctuation energy in soliton models with fermion fields included at both zero and finite temperatures. In our future work we will extend the calculation scheme to the more realistic QCD effective soliton models which include chiral symmetry to study the properties of nucleons in the thermal medium.

The organization of this paper is as follows: in section II the level summation calculation of vacuum energy of the soliton in FL model is performed at the zero temperature. In particular we make a clear illustration of how to do the renormalization. In
section III we show how to calculate the vacuum energy and thermal fluctuation energy of the soliton in the thermal quark medium at finite temperatures. In
section IV, the numerical results are presented and discussed. The last section is the summary and outlook.

\section{Vacuum energies of a soliton and renormalization}
As we are interested in studying the vacuum energy of a fermion system in a nontrivial background of the soliton field, the FL model is suitable for the goal. In previous studies of the FL model most calculations have been carried out in mean field approximation\cite{ref24a,ref24b}. In some studies the quantum corrections are calculated by derivative expansion method or partial wave expansion method in Euclidean Green's function formalism\cite{ref24c,ref24d,ref24e}. However in these calculations the renormalizations are somewhat obscure and have certain ambiguities. In the calculation scheme we have used here all the divergences are removed by the same on-shell mass and coupling constant renormalization scheme which has no ambiguities. And the method used here is generic for the calculation of the vacuum energies of solitons in those soliton models with fermions included in quantum field theory.

We start from the Lagrangian of the FL model as \bea {\cal
L}=\bar\psi(i\gamma_\mu\pl^\mu-g\s)\psi+\012\pl_\mu\s\pl^\mu\s-U(\s),
\eea where\bea U(\s)=\01{2!}a\s^2+\01{3!}b\s^3+\01{4!}c\s^4+B.
\eea $\p$ and $\s$ are the quark field and the
phenomenological scalar field. The model parameters $a, b, c$, and $g$ are generally fitted to confront the properties
of the nucleon in vacuum. $B$ is the bag constant. In the mean field
approximation the sigma field is treated as the classical field
and certain occupied valence levels for the quarks are included.
For a static background $\s$ field and $N$ valence quarks we have
\bea \s(\vec r,t)=\s(\vec r), \ \ \ \ \ \psi(\vec r,t)=e^{-i\e
t}\sum_{i=1}^{N}\psi_i(\vec r). \eea And the static classical
Euler-Lagrange equations can be derived from the above Lagrangian
as \beq (-i\vec\alpha\cdot\vec\nabla+\beta
g\s)\psi_i=\varepsilon\psi_i, \eeq \beq -\vec\nabla^2\s+\0{d
U(\s)}{d\s}+g\sum\limits_{i=1}^{N}\bar\psi_i\psi_i=0.\label{mean}
\eeq If one further considers the spherical symmetrical configurations
of the fields which means, \bea \s(\vec r)=\s(r), \ \ \ \
\psi_i(\vec r)=\01r\left(
\begin{array}{c}
F(r)\\
i\vec\s\cdot\hat{\vec{r}} G(r)
\end{array}\right)y_{\kappa m},
\eea where $y_{\kappa m}\equiv y_{jm}^{l}$ is the two-component
Pauli spinor harmonic and $\hat{\vec r}$ is the spatial unit vector,
the equations could be written into the spherical coordinate and the forms are, \beq
\left(\0{d}{dr}+\0{\kappa}{r}\right)F(r)+(g\s(r)+\e)G(r)=0,\label{u}
\eeq \beq
\left(\0{d}{dr}-\0{\kappa}{r}\right)G(r)+(g\s(r)-\e)F(r)=0,\label{v}
\eeq \beq \frac{d^2 \sigma (r)}{d r^2}  + \frac{2}{r} \frac{d
\sigma(r)}{d r}-\0{dU(\s)}{d\s} - N g(F^2(r) - G^2(r)) = 0, \label{s} \eeq
where the Dirac quantum number $\kappa=-(l+1)$ and $l$ is the quantum number of the angular momentum. For the study of the nucleon at the mean field
approximation the $N$ valence quarks have been put into the lowest
$s$-wave level which means the angular momentum quantum number
$l=0$ or the Dirac quantum number $\kappa=-1$. The quark wave
functions should satisfy the normalization condition \beq
4\pi\int(F^2(r)+G^2(r))r^2dr=1. \eeq

Usually the potential $U(\s)$ has two
minima corresponding to the two vacuums. For $\s=0$ it is a
local minimum of the potential which is corresponding to the perturbative
vacuum, while for $\s=\s_v$ it is a global minimum corresponding to the
physical vacuum or the true vacuum. In the physical vacuum $\s=\s_v$, the valence quark will acquire a mass as $m_q=g\s_v$. For $N$ valence quarks at the lowest energy level $\e$ there will be a semi-classical soliton solution by numerically solving the above equations. The classical energy of the
soliton could be obtained as \beq E_{cl}=N\e+4\pi\int
drr^2\left[U(\s)+\012\left(\0{d\s}{dr}\right)^2\right]. \label{Ecl}\eeq

Now we go forward to study the quantum fluctuations based on the classical soliton background. The spherical symmetrical soliton solution of the $\s$ field
can be decomposed into the following form, \beq \s(r)=\s_v+
\tilde{\s}(r), \label{bag}\eeq where the first term is the vacuum expectation value of the $\s$ field and the second term is not a quantum fluctuation but a classical part with certain spatial configuration which form is like a Woods-Saxon potential and can be written as
\beq \tilde{\s}(r)=-\0{\s_0}{1+e^{(r-R)/r_0}}. \label{ws}\eeq
Here we only consider the vacuum energy from the Dirac sea, which means only the one loop quantum fluctuation from the sea quarks will be calculated, while the sigma field in our calculation is treated as the background. So the quantum corrections coming from the sigma field will be ignored. The corresponding action for the sea quarks in the background field of the $\tilde{\s}(r)$ is \beq S_{\psi}[\tilde{\s}(r)]=\int d^4x\left [\bar\psi(i\g_\mu\partial^\mu-m_q-g\tilde{\s})\psi+{\cal L}_{ct}\right],\label{spsi}\eeq where the renormalization counterterm is \bea {\cal
L}_{ct}=Z_{\s}\pl_\mu\tilde{\s}\pl^\mu\tilde{\s}-Z_m\tilde{\s}^2-Z_g\tilde{\s}.
\eea
The effective action can be defined as \beq e^{iS_{eff}[\tilde{\s}]}=\0{\int[D\psi][iD\psi^\dag]e^{iS_{\psi}[\tilde{\s}]}}{\int[D\psi][iD\psi^\dag]e^{iS_{\psi}[\tilde{\s}]|_{\tilde{\s}=0}}}. \label{pint} \eeq After integrating out the $\psi$ field the one loop effective action could be written as \beq S_{eff}[\tilde{\s}]=Tr\log\0{D[\tilde{\s}]}{D[\tilde{\s}]|_{\tilde{\s}=0}}+S_{ct}, \label{seff} \eeq where $D$ is the Dirac operator which form is $D=i\g_\mu\partial^\mu-m_q-g\tilde{\s}$ and $S_{ct}$ is the counterterm. The effective one loop vacuum energy could be derived by $E_{vac}=-S_{eff}/\int dt$ and the result is \beq E_{vac}[\tilde{\s}]=E_{vac}^{\psi}[\tilde\s]+E_{ct}[\tilde\s], \label{Eeef} \eeq where $E_{ct}$ is the counterterm and $E_{vac}^{\psi}$ is the difference of energy of a filled negative energy Dirac sea with the background field $\tilde\s(r)$ and that of a filled negative energy Dirac sea without the background field which is \beq E_{vac}^{\psi}[\tilde\s]=-\left[\sum_{\alpha} E_{\alpha}-\sum_{k}E_q(k)\right], \label{Evac} \eeq where $E_q(k)=\sqrt{k^2+m_q^2}$. $E_\alpha$ are all the discrete and continuous eigenvalues of the following Dirac equation \beq (-i\vec\alpha\cdot\vec\nabla+\beta m_q+\beta
g\tilde\s)\psi_\alpha=E_\alpha\psi_\alpha. \label{Epsi} \eeq In a large spherical box with the volume going to infinity from Eq.(\ref{Evac}) one can further derive the following form\cite{ref22,ref22a} \beq E_{vac}^{\psi}[\tilde\s]=-\sum\limits_{n}E_n-\sum\limits_{l}(2l+1)\int dk\01\pi\0{d\d_l(k)}{dk}E_q(k), \label{evac1} \eeq where $E_n$ is the possible discrete negative bound energy level of the Dirac equation (\ref{Epsi}), $E_q(k)$ is the negative continuum energy spectrum and $\d_l(k)$ is the scattering phase shift of the quark wave function with angular momentum quantum number $l$. The first derivative of the phase shift to the momentum determines the density of the states in momentum space. However the second term in (\ref{evac1}) is divergent which should be properly renormalized. In order to make it finite one needs a Born subtraction of the phase shift. In one loop order only the first and second Born approximation should be subtracted from the phase shift. Therefore the subtracted phase shift is defined as \beq \bar\d_l(k)\equiv\d_l(k)-\d_l^{(1)}(k)-\d_l^{(2)}(k), \label{deltabar} \eeq in which
$\d_l^{(1)}(k)$ and $\d_l^{(2)}(k)$ are the first and second Born approximations to $\d_l(k)$. All the phase shifts $\d_l(k)$, $\d_l^{(1)}(k)$ and $\d_l^{(2)}(k)$ could be numerically calculated from Eq.(\ref{Epsi}) and it is discussed in detail in Appendix A.
\begin{figure}[tbh]
\begin{center}
\includegraphics[width=180pt,height=70pt]{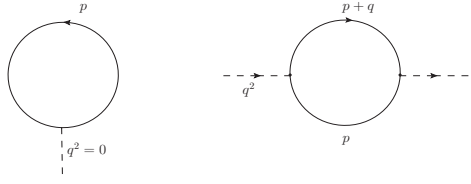}
\end{center}
\caption{One fermion loop diagrams with one and two
insertions.}\label{x01}
\end{figure}

\begin{figure}[tbh]
\begin{center}
\includegraphics[width=205pt,height=110pt]{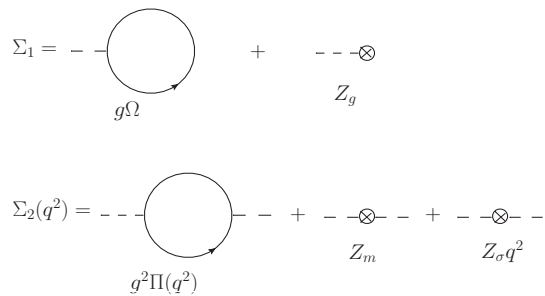}
\end{center}
\caption{The diagrammatic representation of one- and two-point functions arising from the loop and counterterms.}\label{x02}
\end{figure}

In Feynman diagrammatic representation the first and the second Born approximations of the phase shifts are corresponding to the diagrams as one and two insertions of $\s$ fields to the one fermion loop which are shown in Fig.\ref{x01}. The energy terms related to these two Feynman diagrams should be properly renormalized and it will yield a finite energy correction which should be added back to the soliton energy. In order to make the renormalization we define the one- and two-point functions as $\Sigma_1$ and $\Sigma_2(q^2)$, the diagrammatic representation of which are shown in Fig. \ref{x02}. The one-loop diagrams with one insertion and two insertions are denoted by $\O$ and $\Pi(q^2)$ respectively. $Z_g$, $Z_mm_\s^2$ and $Z_\s q^2$ are the corresponding counterterms, which are fixed by choosing the on-shell renormalization conditions as  \bea &&\Sigma_1=g\O+Z_g=0, \nnu \\ &&\Sigma_2(-m_\s^2)=g^2\Pi(-m_\s^2)+Z_mm_\s^2-Z_\s m_\s^2=0, \nnu \\ &&\left.\0{d\Sigma_2(q^2)}{dq^2}\right |_{q^2=-m_\s^2}=g^2\Pi^\prime(-m_\s^2)+Z_\s=0, \label{rn} \eea where $\Pi^\prime(q^2)$ is defined as \beq \Pi^\prime(q^2)\equiv\0{d\Pi(q^2)}{dq^2}. \label{dpi} \eeq The coefficients of the counterterms could be evaluated and the results are \beq Z_g=-g\O, \ \ \ \ Z_\s=-g^2\Pi^\prime(-m_\s^2), \ \ \ \ Z_m=-g^2\left[\0{\Pi(-m_\s^2)}{m_\s^2}+\Pi^\prime(-m_\s^2)\right]. \label{coeffs} \eeq The energy counterterm in Eq.(\ref{Eeef}) in coordinate space could be written as \beq E_{ct}[\tilde\s(x)]=\int d^3x\left\{Z_\s|\nabla\tilde\s(x)|^2+Z_mm_\s^2\tilde\s(x)^2+Z_g\tilde\s(x)\right\}. \label{ect} \eeq The corresponding divergent energy terms which are associated with the Feynman diagrams in Fig.\ref{x01} combined with these counterterms of the energy finally yield the finite piece of the energy written in momentum space as \beq \G_{2}[\tilde\s]=g^2\int \0{q^2dq}{2\pi^2}\Pi_{ren}(q^2)\tilde\s_f(q)^2, \label{gf2} \eeq where the renormalized finite part associated with the second graph in Fig.\ref{x01} is obtained as \beq \Pi_{ren}(q^2)=\Pi(q^2)-q^2\Pi^\prime(-m_\s^2)-\Pi(-m_\s^2)-m_\s^2\Pi^\prime(-m_\s^2), \label{piren} \eeq and $\tilde\s_f(q)$ is the Fourier transform of $\tilde\s(r)$ which is \bea \tilde\s_f(q)&=&\int d^3\vec r\tilde\s_(\vec r)e^{-i\vec q\cdot\vec{r}} \nnu \\ &=&\0{4\pi}{q}\int_0^{\infty} r\sin(qr)\tilde\s(r)dr. \label{fsigma} \eea Note that the energy term associated with the first tadpole diagram in Fig.\ref{x01} has been totally removed by the counterterm. The detail Feynman diagram calculation of $\Pi_{ren}(q^2)$ is presented in Appendix B and the final result is \beq \Pi_{ren}(q^2)=-\0{g^2}{4\pi^2}\left\{\int_0^1dx\left[3x(1-x)q^2+m^2\right]\ln\0{m^2+x(1-x)q^2}{m^2-x(1-x)m^2_{\s}}
+(q^2+m^2_\s)\int_0^1dxx(1-x)\0{3x(1-x)m^2_{\s}-m^2}{m^2-x(1-x)m^2_{\s}}\right\}. \label{renpi} \eeq Substitute this result into Eq.(\ref{gf2}) one can obtain the finite energy correction $\G_2$ from those renormalized Feynman diagrams associated with the Born approximation. Finally the renormalized vacuum energy in the soliton background can be written as \beq
E_{vac}^{ren}[\tilde\s]=-\sum\limits_{n}E_n-\sum\limits_{l}(2l+1)\int dk\01\pi\0{d\bar\d_l(k)}{dk}E_q(k)+\G_2[\tilde\s], \label{evacren} \eeq

\section{Vacuum and thermal fluctuation energies of the soliton at finite temperatures}
In the above section we have obtained the renormalized vacuum energy in the background of the soliton at zero temperature. As it is important to consider the possible medium effects to the hadrons in some real cases, theoretically one should discuss thermal quantum corrections of the soliton from the vacuum and medium. In this section we will discuss the vacuum energy and the thermal fluctuation energy in the background of the soliton at finite temperatures. The physical picture in our discussion is that a soliton is embedded in a thermal quark medium where the negative energy spectrum are deformed and the sea quarks could be excited at a finite temperature while keeping a zero net quark density. The temperature plays its roles in two aspects in the calculation of the vacuum and the thermal fluctuation energies. Firstly at a mean field level the spatial configuration of the soliton will vary with a varying temperature and can be determined at a given temperature. With the given soliton background at the certain temperature the vacuum and the thermal fluctuation energies can be further calculated. Secondly in calculating the fluctuation of the one loop diagram at finite temperatures the evaluating technics of the imaginary time formalism will be used, which means after the Matsubara frequency sum the one loop fluctuation energy is divided into two parts. One is the vacuum part which form in expression is the same as the vacuum energy at zero temperature except for the different soliton background at the finite temperature. It is also called zero point energy of the vacuum which is implicitly temperature dependent. The other part is the thermal fluctuation energy which is explicitly temperature dependent with thermal distribution functions in its form. Only the zero point energy part at finite temperatures need to be renormalized and its process is as the same as that of the zero temperature case except for the temperature dependent soliton background. The explicit finite temperature part of the thermal fluctuation energy is not divergent and needs not to be renormalized.

First we solve the soliton equations at finite temperatures in the mean field level. The sigma field is replaced by its thermal vacuum expectation value $\s_v^T$ and the thermodynamical potential of the system at finite temperatures is \beq \O(T,\s_v^T)=U(\s_v^T)-\nu_qT\int\0{k^2dk}{\pi^2}\ln\left[1+e^{- E_q^T(k)/T}\right], \label{omgT} \eeq where $\nu_q$ is the degenerate factor of the quarks, $E_q^T(k)=\sqrt{k^2+{m_q^T}^2}$ and $m_q^T=g{\s_v^T}$. In order to obtain the soliton at finite temperature one can replace the potential $U(\s)$ by the thermodynamical potential $\O(T,\s_v^T)$ in solving the soliton Eqs.(\ref{u}),(\ref{v}) and (\ref{s}). Then the classical solitons could be solved at different temperatures\cite{ref24}.

Second we calculate the fluctuation of the one loop diagram in the soliton background at finite temperatures. From Eq.(\ref{seff}), after evaluating the trace $logD/D_0$ in imaginary time formalism one can obtain the following expression of the vacuum energy and the thermal fluctuation energy in the background of the soliton at finite temperatures, \beq E_{vac}^{T}[\tilde{\s}_T]=-\left[\sum_{\alpha} E_{\alpha}^T-\sum_{k}E_q^T(k)\right]+E_{ct}[\tilde\s_T], \label{EvacT} \eeq \beq E_{fluc}^{T}[\tilde\s_T]=\sum_{\g} \0{E_{\g}^T}{e^{E^T_\g/T}+1}-\sum_{k}\0{E_q^T(k)}{e^{E_q^T(k)/T}+1}, \label{Etherm} \eeq where $\tilde\s_T$ is the soliton background at the finite temperature and one should notice that at finite temperature the soliton field is written as \beq \s_T(r)=\s_v^T+\tilde{\s}_T(r), \label{sgmT}\eeq where $\s_v^T$ is the vacuum expectation value of the $\s$ field at finite temperatures. $E_\alpha^T$ and $E_\g^T$ are the possible eigenvalues of the Dirac equation (\ref{Epsi}) with a thermal quark mass $m_q^T$ in the soliton background $\tilde{\s}_T(r)$ and $E_q^T(k)$ is continuous free energy spectrum without the soliton background at finite temperatures. The vacuum energy part at finite temperatures is divergent and needs to be renormalized. After doing the same renormalization procedure as that in the zero temperature case except in a different soliton background $\tilde\s_T(r)$ the final renormalized result of the vacuum energy at finite temperatures could be written as \beq
E_{vac}^{T,ren}[\tilde\s_T]=-\sum\limits_{n}E_n[\tilde\s_T]-\sum\limits_{l}(2l+1)E_{ren}^{(l)}[\tilde\s_T]+\G_2[\tilde\s_T], \label{erenT} \eeq where the energy term associated with the angular momentum $l$ is \beq E_{ren}^{(l)}[\tilde\s_T]=\int dk\01\pi\0{d\bar\d_l^T(k)}{dk}E_q^T(k). \label{elT} \eeq $\bar\d_l^T(k)$ is the subtracted phase shift which could be numerically evaluated from Eq.(\ref{Epsi}) in the soliton background $\tilde\s_T$. $\G_2[\tilde\s_T]$ can be determined from Eq.(\ref{gf2}) also in the soliton background $\tilde\s_T$.

With the subtracted phase shift $\bar\d_l^T(k)$ Eq.(\ref{Etherm}) can be further calculated and the final form of the thermal fluctuation energy can be written as \beq
E_{fluc}^{T}[\tilde\s_T]=\sum\limits_{l}(2l+1)E_{T}^{(l)}[\tilde\s_T], \label{eT} \eeq in which the energy term with the thermal distribution function associated with the angular momentum $l$ is \beq E_{T}^{(l)}[\tilde\s_T]=\int dk\01\pi\0{d\bar\d_l^T(k)}{dk}\0{E_q^T(k)}{e^{ E_q^T(k)/T}+1}. \label{eTl} \eeq

\section{Numerical results and discussions}
In this section we will do the numerical evaluations. The parameters of the model are set in the following. As this model was originally devised for studying hadrons especially nucleons, the parameters could be fixed but not uniquely by fitting the nucleon properties. Here we take $N=3$ and choose a set of values of parameters as $a=17.7fm^{-2}, b=-1457.4fm^{-1}, c=20000, g=12.16$\cite{ref25}. The bag constant $B$ is thus fixed by the vacuum value $\s_v$. In the renormalization the mass of the sigma field is set as $m_\s=550MeV$. In the thermodynamical potential the degenerate factor is $\nu_q=12$. In this section we mainly focused on the discussion of the finite temperature vacuum and thermal fluctuation energies of the soliton.
\begin{table}[th]
\caption{\label{t1}The classical soliton energy $E_{cl}$ and the renormalized energy correction $\G_2$ for different temperatures.}
\begin{ruledtabular}
\begin{tabular}{ccccc}
T(MeV)       &  50    & 90   &  110  & 120   \\
\hline
$E_{cl}(fm^{-1})$   & 6.39  & 6.07  & 5.48  & 4.80  \\
$\G_2(fm^{-1})$ & -2.37 & -2.57 & -3.23 & -4.81 \\
\end{tabular}
\end{ruledtabular}
\end{table}

\begin{figure}[tbh]
\begin{center}
\includegraphics[width=210pt,height=150pt]{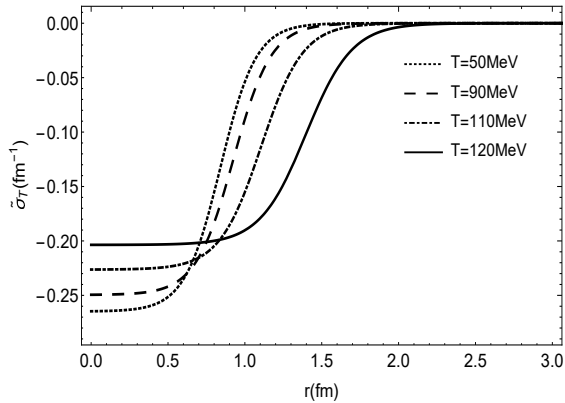}
\end{center}
\caption{The spatial profiles of $\tilde\s_T$ at different temperatures. The dotted, dashed, dot-dashed
and solid lines are for the cases of $T=50,90,110,120MeV$ respectively.}\label{x1}
\end{figure}

From Eqs.(\ref{erenT}) and (\ref{eT}) one could see that the vacuum and thermal fluctuation energies are dependent on the finite temperature soliton background $\tilde\s_T$. As discussed earlier the soliton profiles could be solved at different temperatures in this model. In Fig.\ref{x1} the spatial configurations of solitons $\tilde\s_T$ at different temperatures are plotted. It could be seen that the spatial profile of the soliton is flattened mildly with temperature increasing, which means the soliton size is slowly expanding with the temperature increasing. The first term in the finite temperature vacuum energy (\ref{erenT}) is the possible discrete energy eigenvalue of the Dirac equation (\ref{Epsi}) in the soliton background $\tilde\s_T$. However it turns out that there is no such discrete negative eigenvalue in this system. The third term $\G_2[\tilde\s_T]$ in the vacuum energy expression (\ref{erenT}) is the the correction energy from the renormalization of the Feynman diagrams which has been calculated in Appendix B. At the finite temperature it could be evaluated in the given profile of $\tilde\s_T$. Besides the vacuum energy the classical soliton energy $E_{cl}$ at the different temperature is also evaluated from Eq.(\ref{Ecl}) in the different soliton background $\tilde\s_T$ for comparison. The numerical results of the classical soliton energy and the renormalized correction energy $\G_2$ at different temperatures are shown in table \ref{t1}. One can see that the renormalization correction energy $\G_2$ is negative and its absolute value is increasing with temperature increasing. Meanwhile the classical soliton energy is decreasing with temperature increasing. At certain high temperature the correction energy becomes even comparable with the classical soliton energy.

\begin{figure}[hbt]
\begin{center}
\subfigure{
\begin{minipage}[t]{0.5\linewidth}
\centering
\includegraphics[width=120pt,height=90pt]{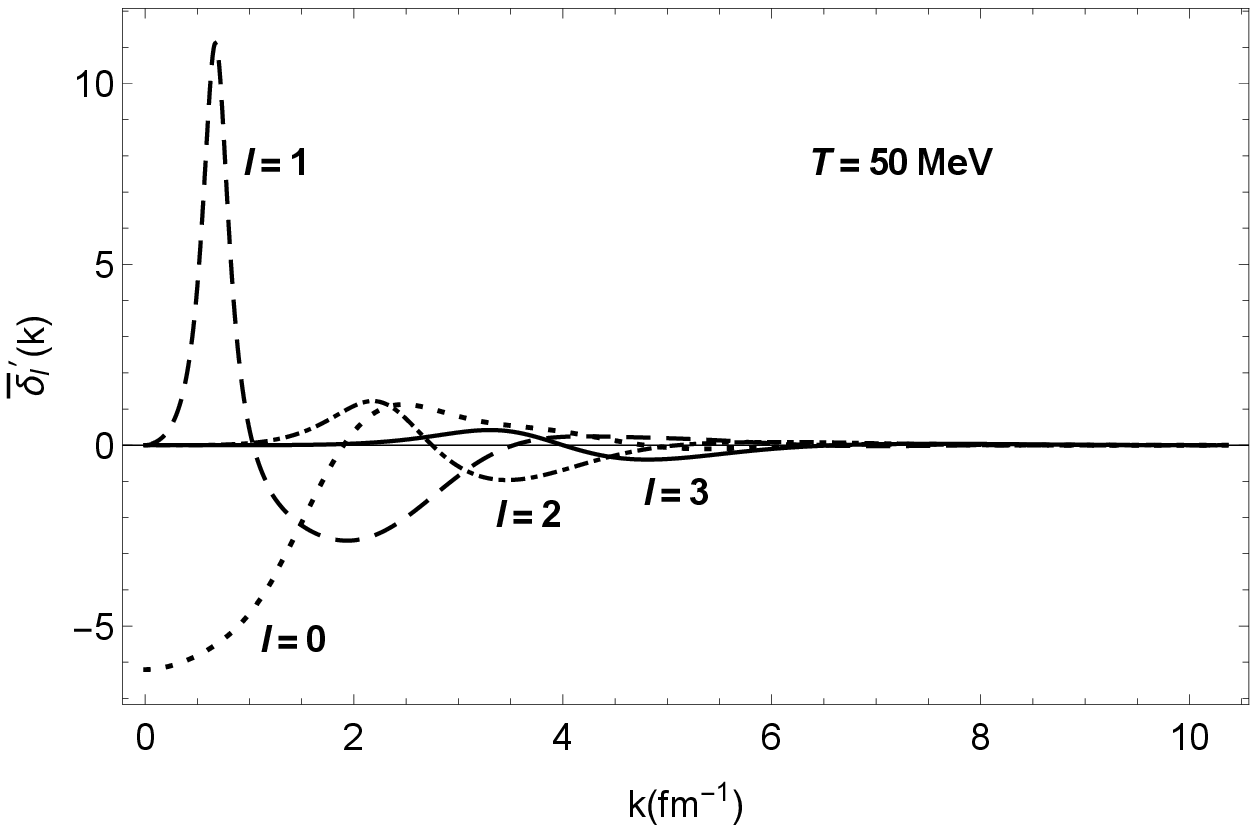}
\includegraphics[width=120pt,height=90pt]{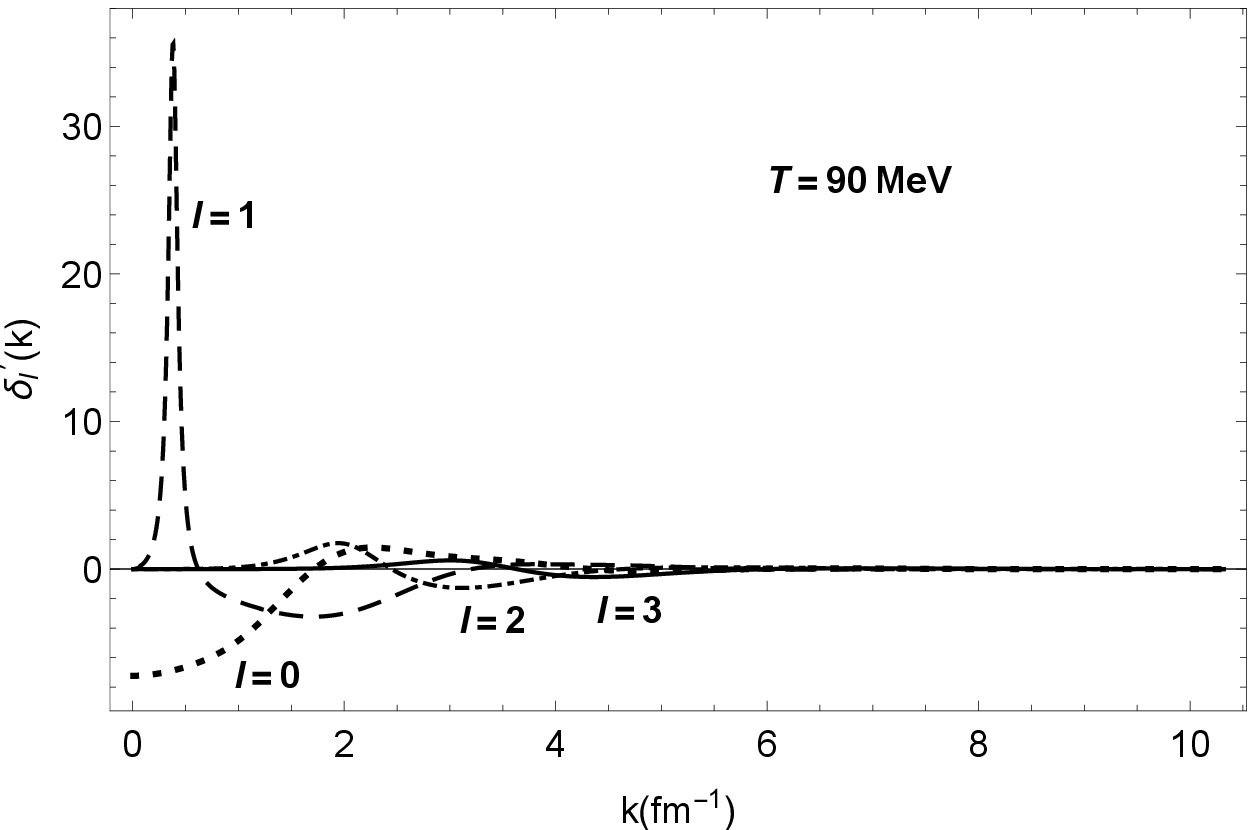}
\end{minipage}
}
\subfigure{
\begin{minipage}[t]{0.5\linewidth}
\centering
\includegraphics[width=120pt,height=90pt]{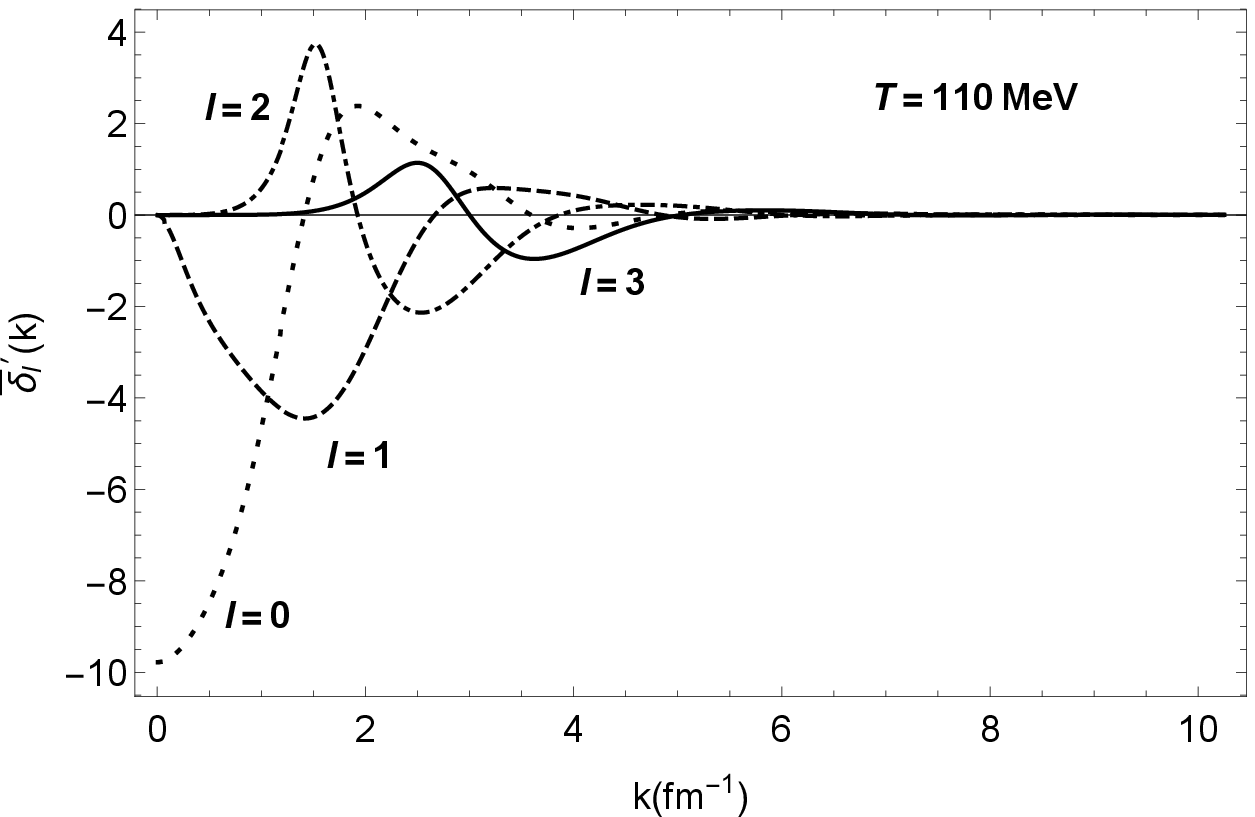}
\includegraphics[width=120pt,height=90pt]{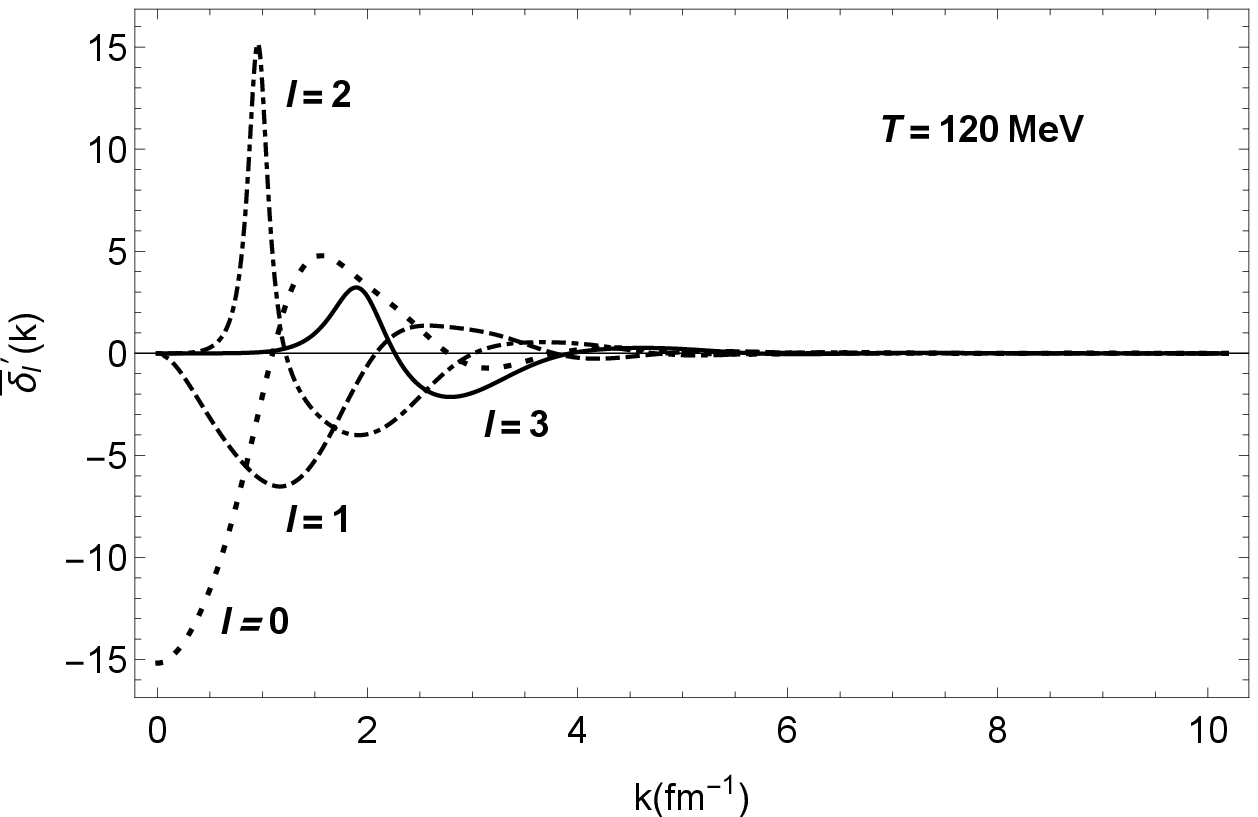}
\end{minipage}
}
\end{center}
\caption{The first order momentum derivative of the subtracted phase
shift $\bar\d_l(k)$ as a function of momentum $k$ for different temperatures. The
dotted, dashed, dot-dashed and solid lines are for the cases of $l=0,1,2,3$
respectively.}\label{x2}
\end{figure}

The second term in the vacuum energy (\ref{erenT}) is the correction energy coming from the integration over the continuous energy eigenvalues. The integration in momentum space in the energy term $E_{ren}^{(l)}$ is dependent on the subtracted phase shift $\bar\delta_l^T$ for different angular momentum $l$. The thermal fluctuation energy of the soliton is decomposed by angular momentum $l$ in Eq.(\ref{eT}). The momentum integration in the energy term (\ref{eTl}) with angular momentum $l$ of the thermal fluctuation energy is also dependent on the phase shift $\bar\delta_l^T$. Based on the finite temperature soliton profiles the subtracted phase shift $\bar\delta_l^T$ could be numerically evaluated. The momentum dependencies of the first order momentum derivatives of the phase shifts $\bar\delta_l^{T}$ for different temperatures are plotted in Fig.\ref{x2}. As we know $l=0,1,2,3$ are quantum numbers of angular momentums corresponding to the s,p,d,f partial waves respectively, in the following discussion we will use s,p,d,f to denote the cases of different quantum number $l$.

One can see the profiles of the first momentum derivative of the phase shifts of p,d,f waves are similar in pattern, while that of the s-wave phase shift is different from them. As the temperature increases, the amplitude of the case of the s-wave phase shift increases slowly, while those of p,d,f-waves increases more remarkably. And one can see that all the profiles of $\bar\d_l(k)$ are squeezed to the left along the momentum $k$ axis with temperature increasing. For example, let us take a close look at the profile of $\bar\d^\prime_l(k)$ for the case $l=1$ namely the p-wave case. When the temperature increases from $T=50MeV$, its amplitude gradually increases and the positions of the positive and negative peaks are moving to the left along the $k$ axis which means the the profile is compressed to the left with temperature increasing. And one can see an interesting result that for the p-wave case as the first positive peak of $\bar\d^\prime_1(k)$ is squeezed to the left along $k$ axis, at certain temperature between $90MeV$ and $110MeV$ it disappears. This leads to a dramatic change of profile of the p-wave $\bar\d^\prime_1(k)$ as temperature increasing from $90MeV$ to $110MeV$. Furthermore it can be seen that at $T=120MeV$ the profile of the d-wave $\bar\d^\prime_2(k)$ becomes similar to the p-wave case at $T=50MeV$. We will make some comments on the interesting results later.

From the evaluated phase shift $\bar\delta_l^{T}$ the momentum integration in the vacuum energy terms $E_{ren}^{(l)}$ and the thermal fluctuation energy terms $E_T^{(l)}$ for different angular momentum $l$ at different temperatures could be further evaluated and the numerical results are presented in tables \ref{t2} and \ref{t3}. First one can see that the vacuum energy term $E_{ren}^{(l)}$ at the finite temperature is $1\sim 6$ orders larger in magnitude than the thermal fluctuation energy part $E_T^{(l)}$ in general. Thus the vacuum energy part $E_{ren}^{(l)}$ gives the main contribution to the correction energy of the soliton at finite temperatures. Second all the values of the vacuum energy terms $E_{ren}^{(l)}$ at different temperatures are negative. However one should notice that in the finite temperature vacuum energy expression (\ref{erenT}) there is a negative sign in front of the vacuum energy term $E_{ren}^{(l)}$. As a result the vacuum energies from different partial waves at finite temperatures are finally positive which means these vacuum energies will increase the total energy of the soliton.

\begin{table}[th]
\caption{\label{t2}The vacuum energy terms $E_{ren}^{(l)}$ for different temperatures.}
\begin{ruledtabular}
\begin{tabular}{ccccc}
T(MeV)                   &  50  & 90    &  110  & 120   \\
\hline
$E_{ren}^{(0)}(fm^{-1})$ & -4.24 & -4.05  & -3.72  & -3.36  \\
$E_{ren}^{(1)}(fm^{-1})$ & -0.54 & -0.57  & -4.8   & -4.29  \\
$E_{ren}^{(2)}(fm^{-1})$ & -0.33 & -0.34  & -0.39  & -0.46  \\
$E_{ren}^{(3)}(fm^{-1})$ & -0.21 & -0.24  & -0.26  & -0.32  \\

\end{tabular}
\end{ruledtabular}
\end{table}

\begin{table}[th]
\caption{\label{t3}The energy terms $E_{T}^{(l)}$ of the thermal fluctuation energy for different temperatures.}
\begin{ruledtabular}
\begin{tabular}{ccccc}
T(MeV)                   &  50  & 90    &  110  & 120   \\
\hline
$E_{T}^{(0)}(fm^{-1})$ & $-4.18\times 10^{-4}$ & $-3.22\times 10^{-2}$  & $-9.84\times 10^{-2}$  & $-1.62\times 10^{-1}$  \\
$E_{T}^{(1)}(fm^{-1})$ & $1.74\times 10^{-4}$ & $1.21\times 10^{-2}$  & $-5.77\times 10^{-2}$  & $-1.08\times 10^{-1}$  \\
$E_{T}^{(2)}(fm^{-1})$ & $7.60\times 10^{-6}$ & $1.83\times 10^{-3}$  & $9.96\times 10^{-3}$  & $2.71\times 10^{-2}$  \\
$E_{T}^{(3)}(fm^{-1})$ & $4.14\times 10^{-7}$ & $2.80\times 10^{-4}$  & $2.47\times 10^{-3}$  & $9.71\times 10^{-3}$  \\

\end{tabular}
\end{ruledtabular}
\end{table}

Let us focus on the numerical results in table \ref{t2}. One can see that the absolute value of the s-wave vacuum energy term $E_{ren}^{(0)}$ is gradually decreasing with temperature increasing, while the absolute values of vacuum energy terms $E_{ren}^{(l)}$ of d,f-waves are gradually increasing with temperature increasing. One can notice that the p-wave vacuum energy term $E_{ren}^{(1)}$ is special. Its absolute value is first increasing then decreasing with temperature increasing. In particular the absolute value of the p-wave vacuum energy increases dramatically when temperature increasing from $90MeV$ to $110MeV$. At relative low temperature ($T\lesssim 90MeV$) the s-wave vacuum energy contribution is dominant and is one order in magnitude larger than the p,d,f-wave vacuum energy. However when temperature is near $T\simeq 110MeV$ the p-wave vacuum energy contribution increases quickly and surpasses the s-wave vacuum energy and becomes the dominant part.

The results here can be qualitatively understood by the view of the partial wave scattering in quantum mechanics. From Fig.\ref{x1} we can see the soliton size is increasing with temperature increasing. The soliton size reflects the spatial size of the scattering center which is an important scale in determining the scattering amplitude or the cross section in scattering problem. The large scattering amplitude will result in a large vacuum energy in our study. When the soliton size is small, we know the s-wave scattering has the largest scattering amplitude or cross section which means the s-wave vacuum energy is dominant compared to those of the other partial waves. When the soliton size becomes larger, the s-wave scattering amplitude will decrease monotonically, while the scattering amplitudes of other partial waves for $l\neq 0$ will increase. Due to the centrifugal potential the $l\neq 0$ partial waves will have resonant scattering. That means at certain soliton size some partial wave for $l\neq 0$ will have maximum scattering amplitude which produces the maximum vacuum energy. As for the p-wave case the resonant scattering greatly enhanced the p-wave scattering amplitude at certain soliton size. This is the reason why the p-wave vacuum energy for certain high temperatures suddenly increases and reaches a maximum value and then decreases with temperature increasing. However in the temperature range of our study the soliton size does not reach the scale to trigger the resonant scattering for the d and f partial waves. Thus the d,f-wave vacuum energies only show the monotonic increasing with temperature increasing. It could be predicted that if the soliton size was increased further, the vacuum energies of the d and f partial waves would reach their maxima sequentially. However it will not happen here as the soliton solution will disappear when the temperature $T\gtrsim 121MeV$ as the result of deconfinement in this model\cite{ref24,ref26}.

Here we can make some comments on the relation between the resonant scattering and the variation results of the phase shifts in Fig.\ref{x2}. As the profile of the p-wave phase shift is squeezed to the left along the $k$ axis, the vacuum energy of p partial wave is slowly increasing. At certain temperatures the first peak in the profile of the p-wave phase shift is squeezed so much to the origin of the $k$ axis and finally it has been pushed out of the range which means the first peak disappears. And this dramatic change of the profile of the p-wave phase shift is a reflection of the resonant scattering of the p partial wave. As a result the vacuum energy of p partial wave suddenly increases and reaches the maximum value. Theoretically the resonant scattering will be there for $l\neq0$ partial waves, so the tendencies of similar variation patterns of the profiles of d and f partial wave phase shifts are also observed in Fig.\ref{x2}. And this is why at $T=120MeV$ the profile of the d-wave phase shift becomes similar to that of the p-wave case at $T=50MeV$. The similar variation patterns to that of the p-wave phase shift will also take place in the cases of d and f-wave phase shifts sequentially if the soliton size would keep increasing.

Finally let us take a look at table \ref{t3}. The absolute values of the thermal fluctuation energies are much smaller than those of the vacuum energies for different partial waves at finite temperatures, because the thermal distribution function as a weight factor in momentum space which value is much less than $1$ has suppressed the result of the integration over the momentum space. The s-wave thermal fluctuation energies are all negative and its absolute value is monotonically increasing with temperature increasing. The s-wave thermal fluctuation energy is the dominant part compared to those energy values of the other partial waves in our temperature range. The p-wave thermal fluctuation energy is first positive and then negative with temperature increasing. Except for the flipping sign the absolute value of p-wave fluctuation energy still keeps increasing with temperature increasing. The thermal fluctuation energies of d and f partial waves are all positive and their values are also monotonically increasing with temperature increasing. So it could be seen that the absolute values of the thermal fluctuation energies for all the partial waves are monotonically increasing with temperature increasing.

\section{Summary and outlook}
In summary we have performed the calculation of the one loop quantum correction from the vacuum and thermal fluctuations of the soliton in FL model at finite temperatures. All the divergences have been removed by the same on-shell mass and coupling constant renormalization scheme. At finite temperatures it turns out that the quantum correction energy to the soliton from the vacuum is much larger than that from the thermal fluctuation. The correction energies are decomposed to different terms with different values of angular momentum number $l$. For the correction energies from the vacuum, at relatively low temperatures the s-wave energy term is the largest compared to the energy terms of other partial waves. The absolute value of the s-wave energy term decreases with temperature increasing, while the values of the energy terms of other partial waves all increase with temperature increasing. In particular at certain temperature the absolute value of p-wave energy term surpasses the s-wave energy term and becomes the dominant part, which is due to the resonant scattering effect of the p partial wave. For the correction energies from the thermal fluctuations, the absolute values of the thermal fluctuation energies for all the partial waves are monotonically increasing with temperature increasing.

In the present work we mainly focus on developing the calculation scheme for evaluating the vacuum and thermal fluctuation energies of the soliton. In recent years the topological solitons in QCD, such as instantons, dyons and monopoles, have been more systematically studied\cite{ref27,ref28,ref29,ref30}. It is an interesting topic to study the possible nontrivial effect of these topological objects in hadrons especially nucleons in vacuum or medium. In the future study we want to use QCD effective models like the chiral soliton model or the NJL model based on the nontrivial topological QCD vacuum to study the nucleon properties in medium.

\appendix
\section{}
In this appendix we will show how to calculate the phase
shifts $\d_l(k)$, $\d_l^{(1)}(k)$ and $\d_l^{(2)}(k)$. First we rewrite the coupled first order
differential Eqs.(\ref{u}) and (\ref{v}) into the decoupled second order
differential equations as \beq F''-\0{g\s'}{E+g\s}F'-\left[\0\k
r\0{g\s'}{E+g\s}+\0{\k(\k+1)}{r^2}-(E^2-g^2\s^2)\right]F=0,
\label{F} \eeq \beq G''+\0{g\s'}{E-g\s}G'-\left[\0\k
r\0{g\s'}{E-g\s}+\0{\k(\k-1)}{r^2}-(E^2-g^2\s^2)\right]G=0,
\label{G} \eeq where the prime denotes the differentiation with
respect to $r$ and $\s=\s_v+\tilde{\s}$. For the evaluation of the phase shift both equations will give the same result. In the following discussion Eq.(\ref{F}) will be used for the calculation of the phase shift. When $r>>R$ the asymptotic form of Eq.(\ref{F}) is
\beq F''-\left[\0{\k(\k+1)}{r^2}-k^2\right]F=0, \eeq where
$k^2=E^2-g^2\s^2_v$. The solutions will be spherical Hankel
functions. Meanwhile it should satisfy that $F(r)\to 0$ with $r\to 0$. There are two linearly independent solutions
\beq F^{(1)}_l(r)=e^{i\b_l(k,r)}rh^{(1)}_l(kr), \eeq \beq
F^{(2)}_l(r)=e^{-i\b^*_l(k,r)}rh^{(2)}_l(kr), \eeq where
$h^{(1)}_l(kr)$ and $h^{(2)}_l(kr)$ are the Hankel functions of
the first and second kinds and $h^{(2)}_l(kr)=h^{(1)*}_l(kr)$. The
function $\b_l(k,r)$ should satisfy $\b_l(k,r)\to 0$ as $r\to
\infty$. Then the scattering solution is \beq
F_l(r)=F^{(2)}_l(r)+e^{i\d_l(k)}F^{(1)}_l(r), \eeq and obeys
$F_l(0)=0$, which leads to the result of the scattering phase
shift \beq \d_l(k)=-2\textmd{Re}\b_l(k,0), \eeq where
$\textmd{Re}$ means the real part. It is obvious that the phase shift could be evaluated from $\b_l$. By substituting $F^{(1)}_l$
into Eq.(\ref{F}) one could obtain the equation of $\b_l$
\beq i\b''_lrh_l+2i\b'_l(h_l+rh'_l)-\b^{\prime
2}_lrh_l-\0{g\s'}{E+g\s}(i\b'_lrh_l+h_l+rh'_l)-\left[\0{\k}{r}\0{g\s'}{E+g\s}+g^2(\s^2-\s^2_v)\right]rh_l=0.
\label{ricatti} \eeq In the fixed background soliton field of
$\s(r)$ this equation could be numerically solved to obtain the
phase shift $\d_l(k)$.

To get the Born approximation to the phase shift one should expand
$\b_l$ in powers of $g$ as \beq \b_l=g\b_{l1}+g^2\b_{l2}+\cdots.
\label{beta} \eeq Substituting the expansion (\ref{beta}) into
Eq.(\ref{ricatti}) and neglecting the higher order terms
$O(g^3)$ one can obtain a set of coupled differential equations
about $\b_{l1}$ and $\b_{l2}$ as \beq
i\b''_{l1}rh_l+(2i\b'_{l1}-\0{\s'}{E})(h_l+rh'_l)-\0{\k\s'}{E}h_l=0,
\eeq \beq i\b''_{l2}rh_l-\b^{\prime
2}_{l1}rh_l-\0{i\s'}{E}\b'_{l1}rh_l+
(2i\b'_{l2}+\0{\s'\s}{E^2})(h_l+rh'_l)+\left[\0{\k\s'\s}{rE^2}-(\s^2-\s^2_v)\right]rh_l=0.
\eeq These equations could be numerically solved to obtain the
first and second Born approximations of the phase shift namely
$\d^{(1)}_l$ and $\d^{(2)}_l$ as \beq
\d^{(1)}_l=-2g\textmd{Re}\b_{l1}(k,r=0), \ \ \
\d^{(2)}_l=-2g^2\textmd{Re}\b_{l2}(k,r=0). \eeq
\section{}
In this appendix we will show the Feynman diagram calculation of $\Pi_{ren}(q^2)$. The second diagram in
Fig.\ref{x01} can be evaluated in the following form \bea
i\Pi(q^2)&=&-g^2\int\0{d^4p}{(2\pi)^4}\textmd{Tr}\left[S(\slashed{p}+\slashed{q})S(\slashed{p})\right]
\\ &=&-4g^2\int\0{d^4p}{(2\pi)^4}\0{N}{[(p+q)^2+m^2][p^2+m^2]}, \eea
where $N=(p+q)p+m^2$. Combining the two denominators using the
Feynman parameter one has \beq
\01{(p+q)^2+m^2}\01{p^2+m^2}=\int_0^1dx\0{1}{(k^2+D)^2}, \eeq
where $k=p+xq$ and $D=x(1-x)q^2+m^2$. Changing the integration
variable $p\to k$ one gets \beq
i\Pi(q^2)=-4g^2\int\0{d^4k}{(2\pi)^4}\int_0^1dx\0{N}{(k^2+D)^2},
\eeq where $N=k^2-x(1-x)q^2+m^2+(1-2x)kq$. The divergent integral
could be regulated by dimensional regulation and using the
following formulas \bea
\int\0{d^dq}{(2\pi)^d}\0{(q^2)^a}{(q^2+D)^b}=\0{\G(b-a-\012d)\G(a+\012d)}{(4\pi)^{d/2}\G(b)\G(\012d)}D^{-(b-a-d/2},
\\
\G(-n+x)=\0{(-1)^n}{n!}\left[\01x-\g+\sum\limits_{k=1}^{n}k^{-1}+O(x)\right],
\eea where $d=4-\e$. The regulated result is \beq
\Pi(q^2)=\0{g^2}{4\pi^2}\left\{\01{\e}(q^2+2m^2)+\016q^2+m^2-\int_0^1dx\left[3x(1-x)q^2+m^2\right]\ln D\right\}. \eeq Substituting this expression of $\Pi$ into Eq.(\ref{piren}) the divergent part will be canceled and one obtains the finite result as shown in Eq.(\ref{renpi}).

\begin{acknowledgments}
The author S. Shu is very thankful to Edward Shuryak for the helpful
discussions and suggestions for the series of works in this research direction.
\end{acknowledgments}

\end{document}